\documentclass[aps,showpacs,nofootinbib]{revtex4}

\usepackage{graphicx}
\usepackage{graphics}
\usepackage{amssymb}
\usepackage{amsmath}

\newcommand{\half}{\textstyle{\frac{1}{2}}}

\newcommand{\be}{\begin{equation}}
\newcommand{\ee}{\end{equation}}
\newcommand{\bea}{\begin{eqnarray}}
\newcommand{\eea}{\end{eqnarray}}
\newcommand{\beal}{\begin{align}}
\newcommand{\eal}{\end{align}}
\newcommand{\bespl}{\begin{split}}
\newcommand{\espl}{\end{split}}
\newcommand{\nn}{\nonumber}
\newcommand{\nslash}{\kern 0.2 em n\kern -0.50em /}
\newcommand{\kslash}{\kern 0.2 em k\kern -0.45em /}
\newcommand{\pslash}{\kern 0.2 em p\kern -0.50em /}
\newcommand{\Sslash}{\kern 0.2 em S\kern -0.50em /}
\newcommand{\Pslash}{\kern 0.2 em P\kern -0.50em /}
\newcommand{\Rslash}{\kern 0.2 em R\kern -0.50em /}
\newcommand{\open}{{<\kern -0.3 em{\scriptscriptstyle )}}}

\begin{document}
\title{Monte Carlo simulation of events with Drell-Yan lepton pairs from 
antiproton-proton collisions: the fully polarized case}

\author{A.~Bianconi}
\email{andrea.bianconi@bs.infn.it}
\affiliation{Dipartimento di Chimica e Fisica per l'Ingegneria e per i 
Materiali, Universit\`a di Brescia, I-25123 Brescia, Italy, and \\
Istituto Nazionale di Fisica Nucleare, Sezione di Pavia, I-27100 Pavia, Italy}

\author{Marco Radici}
\email{marco.radici@pv.infn.it}
\affiliation{Dipartimento di Fisica Nucleare e Teorica, Universit\`{a} di 
Pavia, and\\
Istituto Nazionale di Fisica Nucleare, Sezione di Pavia, I-27100 Pavia, Italy}

\begin{abstract}
In this paper, we extend the study of Drell-Yan processes with antiproton beams
already presented in a previous work. We consider the fully polarized
$\bar{p}^\uparrow p^\uparrow \rightarrow \mu^+ \mu^- X$ process, because this is
the simplest scenario for extracting the transverse spin distribution of quarks,
or transversity, which is the missing piece to complete the knowledge of the
nucleon spin structure at leading twist. We perform Monte Carlo simulations for
transversely polarized antiproton and proton beams colliding at a center-of-mass 
energy of interest for the future HESR at GSI. The goal is to possibly establish 
feasibility conditions for an unambiguous extraction of the transversity from 
data on double spin asymmetries.
\end{abstract}

\pacs{13.85.-t,13.85.Qk,13.88+e} 

\maketitle


\section{Introduction}
\label{sec:intro}

Building the nonperturbative structure of the nucleon bound state in QCD requires,
first of all, the knowledge of the leading (spin) structure of the nucleon in
terms of quarks and gluons. The transverse spin distribution (in jargon,
transversity) constitutes the missing cornerstone that completes such knowledge 
together with the well known and measured unpolarized and helicity 
distributions~\cite{Artru:1990zv,Jaffe:1991kp,Jaffe:1996zw,Barone:2003fy}. 

From the technical point of view, the transversity is not diagonal in the parton
helicity basis, hence the jargon of chiral-odd function. But in the transverse
spin basis it is diagonal and it can be given the probabilistic interpretation of 
the mismatch between the numbers of partons with spin parallel or antiparallel to
the transverse polarization of the parent hadron. In a nucleon (and, in general,
for all hadrons with spin $\half$), the gluon has no transversity because of the
mismatch in the change of helicity units; hence, the evolution of transversity for
quarks decouples from radiative gluons. But it also decouples from charge-even $q
\bar{q}$ configurations of the Dirac sea, because it is odd also under charge
conjugation transformations. In conclusion, the transversity should behave like a
nonsinglet function, describing the distribution of a valence quark spoiled by any
radiative contribution~\cite{Jaffe:1996zw}. The prediction of a weaker evolution 
of transversity with respect to the helicity distribution is counterintuitive and 
it represents a basic test of QCD in the nonperturbative domain. 

The first moment of transversity is related to the chiral-odd twist-2 tensor 
operator $\sigma^{\mu\nu} \gamma_5$, which is not part of the hadron full angular
momentum tensor~\cite{Jaffe:1996zw}. Therefore, the transversity is not related 
to some partonic fraction of the nucleon spin, but it opens the door to studies 
of chiral-odd QCD operators and, more generally, of the role of chiral symmetry 
breaking in the nucleon structure. In particular, another basic test of QCD 
should be possible, namely to verify the prediction that the nucleon tensor 
charge is much larger than its helicity, as it emerges from preliminary lattice 
studies~\cite{Aoki:1996pi}. 

From the experimental point of view, the transversity is quite an elusive object:
being chiral-odd, it needs to be coupled to a chiral-odd partner inside the cross
section. As such, it is systematically suppressed in inclusive Deep-Inelastic
Scattering (DIS)~\cite{Jaffe:1993xb}. The first pioneering work about the 
strategy for its measurement suggested the production of Drell-Yan lepton pairs 
from the collision of two transversely polarized proton 
beams~\cite{Ralston:1979ys}. By flipping the spin of one of the two beams,
it is possible to build a double spin asymmetry in the azimuthal angle of the
final pair, that displays at leading twist the factorized product of the two 
transversities for the colliding quark-antiquark pair. This is the simplest
possible scenario, since no other unknown functions are involved. But, in
principle, the transverse spin distribution of an antiquark in a transversely
polarized proton cannot be large. Moreover, the combined effect of evolution and 
of the Soffer inequality seems to constrain the double spin asymmetry to very 
small values~\cite{Martin:1998rz,Barone:1997mj}. 

Alternatively, in semi-inclusive reactions the transversity can appear in the
leading-twist part of the cross section together with a suitable chiral-odd
fragmentation function~\cite{Mulders:1996dh}. For 1-pion inclusive production, 
like in $p p^\uparrow \rightarrow \pi X$ or $e p^\uparrow \rightarrow e' \pi X$ 
reactions, the chiral-odd partner can be identified with the Collins 
function~\cite{Collins:1993kk}. However, the situation is not so clear, since 
other competitive mechanisms (like, e.g., the Sivers effect~\cite{Sivers:1990cc}) 
can produce the same single spin asymmetry when flipping the spin of the 
transversely polarized target. This happens because one crucial requirement is 
that the spin asymmetry must keep memory of the transverse momentum of the 
detected pion with respect to the jet axis, and, consequently, of the intrinsic
transverse momentum of the parton. Several nonperturbative mechanisms can be
advocated to relate the latter to the transverse polarization (see, among others,
the Refs.~\cite{Gamberg:2003eg,Bacchetta:2003rz,Efremov:2003tf,Anselmino:2004ky}), 
ultimately involving the orbital angular motion of partons inside 
hadrons~\cite{Brodsky:2002cx,Belitsky:2002sm,Burkardt:2003je}. 

Rapid developments are emerging in this field. In particular, new azimuthal 
asymmetries are being deviced to extract the transversity at leading twist while,
at the same time, circumventing the problem of an explicit dependence upon the
transverse momentum. When the 2-pion inclusive production~\cite{Collins:1994ax} 
is considered both in hadronic collisions~\cite{Bacchetta:2004it} and lepton DIS 
with a transversely polarized target~\cite{Bacchetta:2002ux,Bacchetta:2003vn}, the
chiral-odd partner of transversity is represented by the interference 
fragmentation function $H_1^{\open}$~\cite{Bianconi:1999cd}, of which a specific
momentum enters the leading-twist single spin asymmetry depending only upon the
total momentum and invariant mass of the pion 
pair~\cite{Radici:2001na,Jaffe:1998hf}. 

Experimentally, some recent measurements of semi-inclusive reactions with 
hadronic~\cite{Bravar:2000ti} and leptonic~\cite{Airapetian:2004tw} beams have 
been performed using pure transversely polarized proton targets. New experiments 
are planned in several laboratories (HERMES at DESY, CLAS at TJNAF, COMPASS at 
CERN, RHIC at BNL). In particular, we mention the new project of an antiproton 
factory at GSI in the socalled High Energy Storage Ring (HESR). In fact, the 
option of having collisions of (transversely polarized) proton and antiproton 
beams should make it possible to study single and double spin asymmetries in 
Drell-Yan 
processes~\cite{pax,pax2,assia,Maggiora:2005cr,Efremov:2004qs,Anselmino:2004ki} 
with the further advantage of involving unsuppressed distributions of valence 
partons, like the transversely polarized antiquark in a transversely polarized 
antiproton. 

In a previous paper~\cite{Bianconi:2004wu}, we have explored the Drell-Yan 
processes $\bar{p}\,p^{(\uparrow)} \rightarrow \mu^+\,\mu^-\,X$. For the
single-polarized one, in the leading-twist single spin asymmetry the transversity 
happens convoluted with another chiral-odd function~\cite{Boer:1999mm}, which is 
likely to be responsible for the well known (and yet unexplained) violation of 
the Lam-Tung sum rule, an anomalous azimuthal asymmetry in the corresponding 
unpolarized cross section~\cite{Falciano:1986wk,Guanziroli:1987rp,Conway:1989fs}. 
Monte Carlo simulations have been performed for several kinematic configurations 
of interest for HESR at GSI, in order to estimate the minimum number of events 
needed to unambiguously extract the above chiral-odd distributions from a 
combined analysis of the two asymmetries. 

In this paper, we will extend that work by considering numerical simulations for 
the fully polarized Drell-Yan process $\bar{p}^\uparrow \, p^\uparrow \rightarrow 
\mu^+ \, \mu^- \, X$, again at several kinematic configurations of interest for 
the HESR project at GSI. Since this is the simplest possible combination at 
leading twist involving, moreover, the dominant valence contribution of the 
transversity, the goal is to possibly establish feasibility conditions for its  
unambiguous direct extraction from data on double spin asymmetries. Most of the 
details of the simulation have been presented in Ref.~\cite{Bianconi:2004wu} and 
they will be briefly reviewed in Sec.~\ref{sec:kin}, together with a description 
of the kinematics. Results are discussed in Sec.~\ref{sec:out} and some final 
conclusions are drawn in Sec.~\ref{sec:end}.


\begin{figure}[h]
\centering
\includegraphics[width=7cm]{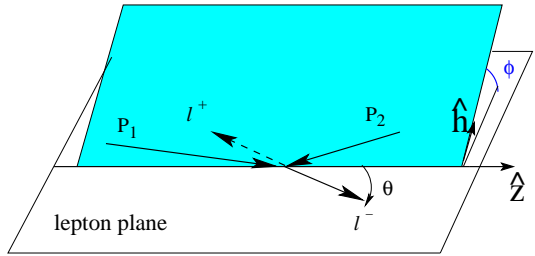}
\caption{The Collins-Soper frame.}
\label{fig:dyframe}
\end{figure}

\section{Theoretical framework and numerical simulations}
\label{sec:kin}

In a Drell-Yan process, a lepton with momentum $k_1$ and an antilepton with
momentum $k_2$ (with $k_{1(2)}^2 \sim 0$) are produced from the collision of two
hadrons with momentum $P_1$, mass $M_1$, spin $S_1$, and $P_2, M_2, S_2$,
respectively (with $P_{1(2)}^2=M_{1(2)}^2, \; S_{1(2)}^2=-1, \; P_{1(2)}\cdot
S_{1(2)}=0$). The center-of-mass (cm) square energy available is $s=(P_1+P_2)^2$ 
and the invariant mass of the final lepton pair is given by the time-like momentum
transfer $q^2 \equiv M^2 = (k_1 + k_2)^2$. In the kinematical regime where $M^2,s
\rightarrow \infty$, while keeping the ratio $0\leq \tau = M^2/s \leq 1$ limited,
the lepton pair can be assumed to be produced from the elementary annihilation of
a parton and an antiparton with momenta $p_1$ and $p_2$, respectively. If $P_1^+$
and $P_2^-$ are the dominant light-cone components of hadron momenta in this
regime, then the partons are approximately collinear with the parent hadrons and
carry the light-cone momentum fractions $0\leq x_1 = p_1^+ / P_1^+ , \; x_2 =
p_2^- / P_2^- \leq 1$, with $q^+ = p_1^+, \; q^- = p_2^-$ by momentum
conservation~\cite{Boer:1999mm}. It is usually convenient to study the problem in 
the socalled Collins-Soper frame~\cite{Collins:1977iv} (see 
Fig.~\ref{fig:dyframe}), where
\bea
\hat{t} &= &\frac{q}{\sqrt{q^2}} \nn \\
\hat{z} &= &\frac{x_1 P_1}{\sqrt{q^2}} - \frac{x_2 P_2}{\sqrt{q^2}} \nn \\
\hat{h} &= &\frac{q_{_T}}{|{\bf q}_{_T}|} \; ,
\label{eq:colsop-frame}
\eea
and ${\bf q}_{_T}$ is the transverse momentum of the final lepton pair detected 
in the solid angle $(\theta, \phi)$. Azimuthal angles are measured in a plane
perpendicular to $\hat{z}, \hat{t}$, and containing $\hat{h}$.


\begin{figure}[h]
\centering
\includegraphics[width=7cm]{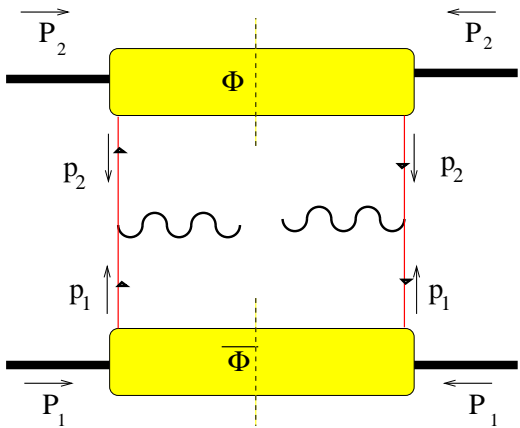}
\caption{The leading-twist contribution to the Drell-Yan process.}
\label{fig:handbag}
\end{figure}

\subsection{Double spin asymmetry}
\label{sec:dsa}

If the invariant mass $M$ is not close to the values of known vector resonances,
under the above mentioned conditions for factorization the elementary annihilation
can be assumed to proceed through a virtual photon converting into the final
lepton pair. Then, the leading-order contribution is represented in
Fig.~\ref{fig:handbag}~\cite{Ralston:1979ys}. The hadronic tensor is given by 
\be
W^{\mu\nu} = \frac{1}{3}\int dp_1^- dp_2^+\, \mbox{Tr} \left[ \bar{\Phi} 
(p_1;P_1,S_1)\, \gamma^\mu\,\Phi(p_2;P_2,S_2) \, \gamma^\nu \right] 
\Bigg\vert_{p_1^+=x_1 P_1^+,\,p_2^- = x_2 P_2^-} \, + \, \left( 
\begin{array}{ccc} q & \leftrightarrow & -q \\ \mu & \leftrightarrow & \nu 
\end{array} \right) \; ,
\label{eq:tensor}
\ee
where the nonlocal correlators for the annihilating antiparton (labeled "1")
and parton (labeled "2") are defined as
\bea
\bar{\Phi} (p_1; P_1,S_1) &= &\int \frac{d^4 z}{(2\pi )^4}\, e^{-i p_1 \cdot z}\, 
\langle P_1 S_1 | \psi(z) \, \bar{\psi}(0) |P_1, S_1 \rangle \; , \nn \\ 
\Phi (p_2; P_2,S_2) &= &\int \frac{d^4 z}{(2\pi )^4}\, e^{i p_2 \cdot z}\, 
\langle P_2, S_2 | \bar{\psi}(0) \, \psi(z) |P_2, S_2 \rangle \; .
\label{eq:phi}
\eea
They correspond to the blobs in Fig.~\ref{fig:handbag} and contain all the soft
mechanisms building up the distribution of the two annihilating partons inside
the corresponding hadrons. 

Inserting into Eq.~(\ref{eq:phi}) the leading-twist parametrization for $\Phi$ and
$\bar{\Phi}$ in terms of the (un)polarized partonic distribution 
functions~\cite{Boer:1998nt}, the Drell-Yan differential cross section for 
transversely polarized hadrons, after integrating upon $d{\bf q}_{_T}$, 
becomes~\cite{Tangerman:1995eh}
\bea
\frac{d\sigma^{\uparrow \uparrow}}{dx_1\, dx_2\, d\Omega} &= &
\frac{\alpha^2}{12 q^2}\, \Bigg[ (1+\cos^2 \theta) \, \sum_f\,e_f^2\, \bar{f}^f_1
(x_1)\,f_1^f(x_2) \nn \\
& &\; + \sin^2 \theta \, \cos 2\phi \, \frac{\tilde{\nu} (x_1, x_2)}{2} \nn \\
& &\; + |{\bf S}_{_{T1}}| \, |{\bf S}_{_{T2}}| \, \sin^2\theta \, 
\cos (2\phi - \phi_{_{S_1}} - \phi_{_{S_2}})\, \sum_f \, e_f^2\, \bar{h}^f_1(x_1)
\, h_1^f(x_2) \nn \\
& &\quad + (1 \leftrightarrow 2) \Bigg] \; ,
\label{eq:cross}
\eea
where $\alpha$ is the fine structure constant, $d\Omega = \sin \theta d\theta
d\phi$, $e_f$ is the charge of the parton with flavor $f$, and $\phi_{_{S_i}}$ is
the azimuthal angle of the transverse spin of hadron $i$ as it is measured with 
respect to the lepton plane in a plane perpendicular to $\hat{z}$ and $\hat{t}$ 
(see Fig.~\ref{fig:dyframe}). The function $f^f_1(x)$ is the usual distribution of
unpolarized partons with flavor $f$, carrying a fraction $x$ of the unpolarized
parent hadron; $h_1^f(x)$ is the transversity for the same flavor and momentum
fraction (analogously for the antiparton distributions). The function 
$\tilde{\nu}$ contains the contribution of the parton distribution 
$h_1^\perp$~\cite{Boer:1999mm}, which describes the influence of the (anti)parton 
transverse polarization on its momentum distribution inside an unpolarized parent 
hadron: it is believed to be responsible for the observed anomalous azimuthal 
asymmetry in the unpolarized Drell-Yan cross section, the socalled violation of 
the Lam-Tung sum rule~\cite{Falciano:1986wk,Guanziroli:1987rp,Conway:1989fs}, 
which no QCD calculation is presently able to justify in a consistent 
way~\cite{Brandenburg:1993cj,Brandenburg:1994wf,Eskola:1994py}.

The double spin asymmetry is defined as 
\bea
A_{_{TT}} &= &\frac{d\sigma^{\uparrow \uparrow} - d\sigma^{\uparrow \downarrow}}
{d\sigma^{\uparrow \uparrow} + d\sigma^{\uparrow \downarrow}} \nn \\
&= &|{\bf S}_{_{T1}}| \, |{\bf S}_{_{T2}}| \, 
\frac{\sin^2\theta}{1+\cos^2\theta}\, \cos (2\phi - \phi_{_{S_1}} - \phi_{_{S_2}})
\, \frac{\sum_f \, e_f^2\, \bar{h}^f_1(x_1) \, h_1^f(x_2) + (1\leftrightarrow 2)}
{\sum_f\,e_f^2\, \bar{f}^f_1(x_1)\,f_1^f(x_2) + (1\leftrightarrow 2)} 
\; .
\label{eq:att}
\eea
In the next Section, we describe the details for numerically simulating 
$A_{_{TT}}$ using the cross section~(\ref{eq:cross}).


\subsection{The Monte Carlo simulation}
\label{sec:mc}

In this Section, we discuss numerical simulations of the double spin asymmetry
$A_{_{TT}}$ of Eq.~(\ref{eq:att}) for the $\bar{p}^\uparrow p^\uparrow \rightarrow
\mu^+ \mu^- X$ process. Our goal is to explore if it is possible to establish
precise conditions in order to determine the feasibility of an unambiguous
extraction of the transversity from data. Most of the technical 
details of the present simulation are mutuated from a previous work, where we
performed a similar analysis for the unpolarized and single-polarized Drell-Yan
process. Therefore, we will heavily refer to Ref.~\cite{Bianconi:2004wu} and 
references therein, in the following. 

As for the kinematics, several options have been considered in 
Ref.~\cite{Bianconi:2004wu}. Since the cross section decreases for increasing 
$\tau$, events statistically tend to accumulate in the phase space part 
corresponding to small $\tau$, where $x_p (x_{\bar{p}})$ fall into the range 
0.1-0.3 dominated by the valence contributions. In fact, the phase space for 
large $\tau = x_p x_{\bar{p}}$ is scarcely populated because the virtual photon 
introduces a $1/M^2 \propto 1/\tau$ factor and the parton distributions become 
negligible for $x_p (x_{\bar{p}}) \rightarrow 1$. Since the spectrum of very low 
invariant masses contains many vector resonances, where the elementary 
annihilation cannot be simply described by a diagram like the one in 
Fig.~\ref{fig:handbag}, it seems more convenient to reach such low values of 
$\tau$ by adequately increasing the cm square energy $s$~\cite{Bianconi:2004wu}. 
Moreover, in this case the elementary annihilation should not be significantly 
affected by higher-order corrections like subleading twists, and the 
leading-twist theoretical framework depicted in Sec.~\ref{sec:dsa} should be 
reliable. Hence, for the HESR at GSI the most convenient setup seems to be the 
option where antiprotons with energy $E_{\bar{p}}$ collide against protons with 
energy $E_p$. Nevertheless, as in Ref.~\cite{Bianconi:2004wu} we will explore 
also the option where the antiproton beam with the same energy hits a fixed 
proton target. 

As for the collider mode, neglecting hadron masses we have 
\be
s \ =\  (P_p + P_{\bar{p}})^2\ \approx\  4\, E_p \, E_{\bar{p}} \; ,
\label{eq:m-coll}
\ee
because for the two colliding beams ${\bf \hat{P}}_p = - {\bf \hat{P}}_{\bar{p}}$.
As in Ref.~\cite{Bianconi:2004wu}, we will select the kinematics where 
antiprotons have energy $E_{\bar{p}} = 15$ GeV and protons $E_p = 3.3$ GeV, such 
that $s\sim 200$ GeV$^2$. If $M$ is constrained in the "safe" range 4-9 GeV 
between the $\bar{c}c$ threshold and the first resonance of the $\Upsilon$ 
family, then $\tau$ falls into the statistically significant range 0.08-0.4 where 
the parton distribution functions are dominated by the valence contribution. At 
the same cm energy, we will consider also the range $1.5 \leq M \leq 2.5$ GeV 
between the $\phi$ and $J/\psi$ resonances, which corresponds to the even lower 
range $0.01 \lesssim \tau \lesssim 0.03$. 

For the fixed target mode, which at present is the selected setup by the PANDA 
collaboration at HESR at GSI~\cite{panda}, we have approximately 
\be
s = (P_p + P_{\bar{p}})^2 \approx 2 \, M_p\, E_{\bar{p}} \; ,
\label{eq:m-fixed}
\ee
such that for the considered $E_{\bar{p}}=15$ GeV it results $s\approx 30$ 
GeV$^2$. For this case, we will restrict the invariant mass to the range $1.5 
\leq M \leq 2.5$ GeV, corresponding to $0.07 \leq \tau \leq 0.2$. In fact, for 
$M > 4$ GeV most events are characterized by large partonic fractional momenta, 
where the quark-antiquark fusion model cannot explain the dimuon production.

The Monte Carlo events have been generated by the following cross 
section~\cite{Bianconi:2004wu}:
\be
\frac{d\sigma}{d\Omega dx_{_F} d\tau d{\bf q}_{_T}} = K \, \frac{1}{s}\, 
A({\bf q}_{_T}, x_{_F}, M)\, F(x_{_F}, \tau) \, \sum_{i=1}^4 \, c_i 
({\bf q}_{_T}, x_{_F},\tau) \, S_i(\theta, \phi, \phi_{_{S_p}},
\phi_{_{S_{\bar{p}}}})\; ,
\label{eq:mc-xsect}
\ee
or, equivalently,
\be
\frac{d\sigma}{d\Omega dx_{\bar{p}} dx_p d{\bf q}_{_T}} = K \, \frac{1}{s}\, 
(x_{\bar{p}}+x_p) \, A({\bf q}_{_T}, x_{\bar{p}}-x_p, M)\,F'(x_{\bar{p}}, x_p)\, 
\sum_{i=1}^4\, c'_i ({\bf q}_{_T}, x_{\bar{p}},x_p) \, S_i(\theta, \phi, 
\phi_{_{S_p}}, \phi_{_{S_{\bar{p}}}}) \; ,
\label{eq:mc-xsect1}
\ee
where the invariant $x_{_F} = x_{\bar p} - x_p$ is the fraction of the available
total longitudinal momentum carried by the two annihilating partons in the 
collision cm frame. As it has been stressed in Ref.~\cite{Bianconi:2004wu}, the 
range of values for $x_{_F}$ depends on the energy. For the considered case of 
$E_{\bar{p}} = 15$ GeV, it results $-0.9 \leq x_{_F} \leq 0.9$, where positive 
values correspond to small $\theta$ angles in the Collins-Soper frame (see 
Fig.~\ref{fig:dyframe}), and viceversa. Equations~(\ref{eq:mc-xsect}) and 
(\ref{eq:mc-xsect1}) imply the approximation of a factorized transverse-momentum 
dependence, which has been achieved by assuming the following phenomenological 
parametrization
\be
A(q_{_T},x_{_F},M) = \frac{5\,\displaystyle{\frac{a}{b}\,\left[ \frac{q_{_T}}{b}
\right]^{a-1}}}{\left[ 1 + \left(\displaystyle{\frac{q_{_T}}{b}}\right)^a 
\right]^6} \; ,
\label{eq:mcqT}
\ee
where $a(x_{_F},M),\,b(x_{_F},M),$ are parametric polynomials given in Appendix A 
of Ref.~\cite{Conway:1989fs} and $q_{_T} = |{\bf q}_{_T}|$. Actually, the 
Drell-Yan events studied in Ref.~\cite{Conway:1989fs} were produced by $\pi - p$ 
collisions; however, the same analysis, repeated for $\bar{p}-p$ 
collisions~\cite{Anassontzis:1987hk}, gives a similar distribution for $q_{_T}$ 
not very close to 0 and not much larger than 3 GeV/c. In addition, the above 
$q_{_T}$ distribution was fitted for $M > 4$ GeV and is singular near $M = 1.5$ 
GeV~\cite{Conway:1989fs}; for $M < 4$ GeV we have assumed the same form with the 
mass parameter in the coefficients $a$ and $b$ fixed to the value $M = 4$ GeV. 

In order to simulate Eq.~(\ref{eq:cross}), the events produced by 
Eqs.~(\ref{eq:mc-xsect}) or (\ref{eq:mc-xsect1}) have been integrated in 
$d{\bf q}_{_T}$. However, apart from theoretical problems related to unwanted 
soft mechanisms, it is anyway not possible to collect events with very small 
$q_{_T}$ because of the collider configuration. Hence, events have been selected 
with $q_{_T} > 1$ GeV/$c$. In our previous work~\cite{Bianconi:2004wu}, the 
searched asymmetry was emphasized by this threshold. Here, no such enhancement 
can be present, of course, because of the further integration. But the
unavoidable dependence in $q_{_T}$, introduced by the lower cutoff, reflects in a
drastic variation of the size of the sample. A precise answer depends on the
experimental setup, but for $q_{_T} > 1$ GeV/$c$ approximately 50\% of the initial
sample is excluded, while for $q_{_T} > 0.5$ GeV/$c$ this fraction of events is 
reduced to 20\%. 

The experimental observation that Drell-Yan pairs are usually distributed with 
$q_{_T} > 1$ GeV/$c$~\cite{Conway:1989fs}, suggests that soft mechanisms are 
suppressed, because confinement induces much smaller quark intrinsic transverse 
momenta, but for the same reason it also indicates sizeable QCD corrections to 
the simple parton model. QCD corrections in the Leading-Log Approximation 
(LLA)~\cite{Altarelli:1979ub} would imply a logarithmic dependence on the scale 
$M^2$ inside the various parameters entering the parton 
distributions~\cite{Buras:1977yj} contained in Eqs.~(\ref{eq:mc-xsect}) and 
(\ref{eq:mc-xsect1}), such that it would determine their DGLAP evolution. 
However, it must be stressed that the key scale is $M$, and its range here 
explored is the same of Refs.~\cite{Conway:1989fs,Anassontzis:1987hk}, where the 
functions $F$ and $c_i$ in Eq.~(\ref{eq:mc-xsect}) [or $F'$ and $c'_i$ in 
Eq.~(\ref{eq:mc-xsect1})] are assumed to be independent of $M$. In particular, 
$F'$ is given by
\be
F'(x_{\bar{p}},x_p) = \frac{\alpha^2}{12 Q^2}\,\sum_f\,e_f^2\,
\bar{f}_1^f(x_{\bar{p}}) \, f_1^f (x_p) + (\bar{p} \leftrightarrow p) \; , 
\label{eq:mcF}
\ee
i.e. it is the azimuthally symmetric unpolarized part of Eq.~(\ref{eq:cross}) 
which has been factorized out. The unpolarized distribution $f_1^f (x)$ for 
various flavors $f=u,d,s$, is parametrized as in Ref.~\cite{Anassontzis:1987hk}. 
QCD corrections in the Next-to-Leading-Log Approximation 
(NLLA)~\cite{Altarelli:1979ub} are responsible for the well known $K$ factor, 
which is roughly independent of $x_{_F}$ and $M^2$ but it grows like 
$\sqrt{\tau}$~\cite{Conway:1989fs}. In accordance with 
Ref.~\cite{Bianconi:2004wu}, for the range of interest $0.08\lesssim \tau 
\lesssim 0.4$ we assume as the best compromise the constant value $K=2.5$. But we 
observe that in an azimuthal asymmetry the corrections to the cross sections in 
the numerator and in the denominator should compensate each other; indeed, the 
smooth dependence of the spin asymmetry on NLLA corrections has been confirmed 
for fully polarized Drell-Yan processes at high cm square 
energies~\cite{Martin:1998rz}. 

The whole solid angle $(\theta, \phi)$ of the final muon pair in the Collins-Soper
frame is randomly distributed in each variable. From Eq.~(\ref{eq:cross}), the 
explicit form of the $q_{_T}$-integrated angular distribution is 
\bea
\sum_{i=1}^4\, c'_i (x_{\bar{p}},x_p) \, S_i(\theta, \phi, \phi_{_{S_p}},
\phi_{_{S_{\bar{p}}}}) &= &1 + \cos^2 \theta \nn \\
& &+ \frac{\nu (x_{\bar{p}},x_p)}{2}\, \sin^2\theta \, \cos 2\phi \nn \\
& &+ |{\bf S}_{_{Tp}}|\,|{\bf S}_{_{T\bar{p}}}| \, c_4 (x_{\bar{p}},x_p)\, 
\sin^2\theta \, \cos (2\phi - \phi_{_{S_p}} - \phi_{_{S_{\bar{p}}}}) \; .
\label{eq:mcS}
\eea
Recalling that the azimuthally symmetric unpolarized part $F'(x_{\bar{p}},x_p)$ of
the cross section has been factorized out, the functions $\nu$ and $c_4$ turn out 
to be
\bea
\nu (x_{\bar{p}},x_p) &= &\frac{\tilde{\nu}(x_{\bar{p}},x_p)}
{\sum_f\,e_f^2\,\bar{f}_1^f(x_{\bar{p}}) \, f_1^f (x_p) + (\bar{p} 
\leftrightarrow p)} \nn \\
c_4 (x_{\bar{p}},x_p) &= &\frac{\sum_f \, e_f^2\, \bar{h}^f_1(x_{\bar{p}})\, 
h_1^f(x_p)+ (\bar{p} \leftrightarrow p)}{\sum_f\,e_f^2\,\bar{f}_1^f(x_{\bar{p}}) 
\, f_1^f (x_p) + (\bar{p} \leftrightarrow p)} \; .
\label{eq:c4nu}
\eea
As for the former, the corresponding azimuthal asymmetry has been studied in 
Ref.~\cite{Bianconi:2004wu} adopting the simple parametrization of 
Ref.~\cite{Boer:1999mm} and testing it against the previous measurement of 
Ref.~\cite{Conway:1989fs}. The latter has been further simplified by assuming that
the contribution of each flavor $f$ to the parton distributions can be 
approximated by a corresponding average function~\cite{Bianconi:2004wu}:
\be
c_4 (x_{\bar{p}},x_p) \sim \frac{\langle \bar{h}_1(x_{\bar{p}})\rangle}
{\langle \bar{f}_1(x_{\bar{p}})\rangle} \, \frac{\langle h_1(x_p) \rangle}
{\langle f_1(x_p) \rangle} \; .
\label{eq:mcc4}
\ee
Four types of analytic dependences will be explored for the ratio 
$\langle h_1(x) \rangle / \langle f_1(x) \rangle$, namely the constants 1 and 
0, and the ascending and descending functions $\sqrt{x}$ and $\sqrt{1-x}$,
respectively. All the functional forms satisfy the Soffer 
bound~\cite{Soffer:1995ww} across the whole range $0\leq x \leq 1$, 
and they show different behaviours in the most relevant range $0.1 \lesssim x
\lesssim 0.4$ for the considered kinematics, as it will be clear in the 
following. The goal is to explore under which conditions such different 
behaviours in the ratio can be recognized also in the corresponding double spin 
asymmetry
\be
A_{_{TT}} = |{\bf S}_{_{Tp}}|\,|{\bf S}_{_{T\bar{p}}}| \, 
\frac{\sin^2\theta}{1+\cos^2\theta}\,
\cos (2\phi - \phi_{_{S_p}} - \phi_{_{S_{\bar{p}}}}) \, c_4 (x_{\bar{p}},x_p) \; ,
\label{eq:mcatt}
\ee
by inserting Eq.~(\ref{eq:mcc4}) into Eq.~(\ref{eq:att}). In fact, in that case 
the measurement of $A_{_{TT}}$ would allow the extraction of unambigous 
information on the analytical form of the tranversity $h_1(x)$. 

The azimuthal asymmetry defined in Eq.~(\ref{eq:att}), i.e. by flipping the
transverse polarization of one of the two beams, can be obtained by changing the
sign of the cosine function in Eq.~(\ref{eq:mcatt}). While in the laboratory 
frame the azimuthal angles of the beam transverse polarization are fixed, in the 
Collins-Soper frame they are variable, since the $\hat{h}$ axis is directed along 
$q_{_T}/|{\bf q}_{_T}|$. Hence, for each randomly distributed 
$\phi_{_{S_{\bar{p}}}}$ two sets of events are accumulated for each bin in $x_p$, 
corresponding to $\phi_{_{S_{\bar{p}}}} = \phi_{_{S_p}}$ (parallel transverse
polarizations, positive cosine function indicated by $U$) and to 
$\phi_{_{S_{\bar{p}}}} = \phi_{_{S_p}} + \pi$ (antiparallel transverse 
polarizations, negative cosine function indicated by $D$). Then, the asymmetry is
constructed as $(U-D)/(U+D)$ and binned in $x_p$ after integrating upon 
${\bf q}_{_T}, x_{\bar{p}}$, and the zenithal angle $\theta$. As in 
Ref.~\cite{Bianconi:2004wu}, the $\theta$ angular distribution is restricted to 
the range 60-120 deg, because the $\sin^2 \theta$ dependence would dilute the 
spin asymmetry if events at small $\theta$ were included. This cutoff produces a 
reduction of events by a factor $\approx 2/5$: larger statistics are obtained at 
the price of smaller absolute sizes of the resulting asymmetry. For small
modifications of the above range, the relative size of statistical error bars and 
asymmetries will not change much. 

A further reduction of the event sample is due to the transverse polarization of
the (anti)proton beams, which is assumed to be 50\% on the average, giving an
overall dilution factor 0.25. In our simulation, this means that an average 75\% 
of events have been sorted assuming no polarization at all, while for 25\% of 
events a full transverse polarization is assumed for both beams. We hope that 
actual polarizations will be larger than 50\%, but at the same time we must be
aware that this fact could be compensated by more realistic parton distributions
that are less close to the Soffer bound than the test functions discussed in this
paper.


\section{Results}
\label{sec:out}

In this section, we present results for the Monte Carlo simulation of double spin
asymmetries for the $\bar{p}^\uparrow \, p^\uparrow \rightarrow \mu^+ \, \mu^- \, 
X$ process in order to explore under which conditions the transversity could be
unambiguously extracted from such data. As explained in the previous section, the
most convenient kinematical option for the HESR at GSI seems, at present, the
collision of an antiproton beam with energy $E_{\bar{p}}=15$ GeV and of a proton
beam with $E_p=3.3$ GeV, such that the available cm square energy is 
$s \approx 200$ GeV$^2$. But we have considered also the option of an antiproton 
beam with the same energy hitting a fixed proton target such that $s\approx 30$ 
GeV$^2$. The overall dilution due to beam transverse polarization is assumed 
0.25. Events are sorted according to the cross section of 
Eq.~(\ref{eq:mc-xsect1}) and supplemented by Eqs.~(\ref{eq:mcF})-(\ref{eq:mcc4}) 
with $|{\bf S}_{_{Tp}}|=|{\bf S}_{_{T\bar{p}}}|=1$ for the 25\% of events, and 0 
for the 75\% of them. 

In the following, we will study the double spin asymmetry of Eq.~(\ref{eq:mcatt})
generated by the $\cos (2\phi - \phi_{_{S_{\bar{p}}}} - \phi_{_{S_p}})$ dependence
of Eq.~(\ref{eq:mcS}) in the Collins-Soper frame: positive values of the cosine 
function $(U)$ correspond to parallel transverse polarizations of the two beams, 
negative values $(D)$ to antiparallel polarizations. The asymmetry $(U-D)/(U+D)$
is then constructed for each bin $x_p$ by integrating upon the other variables 
${\bf q}_{_T}, x_{\bar{p}}, \theta$. In some cases we also show the 
corresponding bidimensional $(x_p, x_{\bar{p}})$ distributions. 

Each displayed histogram contains 17000 events. According to the
reduction factors due to the cuts in $M$, $q_{_T}$, and $\theta$ (as discussed in 
the previous section), we need to consider an initial sample of 80000 events, 
which become 40000 by applying the cutoff in $q_{_T}$, and are further reduced to
the final 17000 after the cutoff in $\theta$. The discussed slightly smoother cut
$q_{_T} > 0.5$ GeV/$c$ implies a starting sample of 68000 events to arrive at the
same final 17000 events of the histogram. The $q_{_T}$ distribution of
Eq.~(\ref{eq:mcqT}) is phenomenological, therefore these reduction factors should 
be realistic (at least in the mass range $4 \leq M \leq 9$ GeV). In our Monte 
Carlo simulation it is not possible to predict how many muon pairs would be 
produced with no mass cuts at all, since we take into account physical devices 
that dominate in certain mass ranges only. 

The asymmetry $(U-D)/(U+D)$ will be calculated only for those $x_p$ bins that are
statistically significant, namely that contain a minimum number of events in order
to avoid large fluctuations of purely statistical origin. For the bidimensional
$(x_p, x_{\bar{p}})$ distribution, the cutoff is of at least 10 events per bin.
For the $x_{\bar{p}}$-integrated distribution, the cutoff is of at least 100 
events per bin. Anyway, these statistical cuts do not affect much the overall 
number of surviving events. This latter sample will be referred to as the sample
of "good" events, in the sense that it contains all the events surviving all the
described cuts. As already anticipated, in the histograms of the following figures
the "good" events amount to 17000. In Ref.~\cite{Bianconi:2004wu}, we already
discussed the relation between the number of "good" events and the running time
for an experiment at a given machine luminosity. The same conclusions apply here,
so that a sample of 10000-30000 events seems a reasonable estimate for a sensible
measurement, as it will be clear in the following.

As in the previous paper~\cite{Bianconi:2004wu}, statistical errors are obtained 
by making 20 independent repetitions of the simulation for each considered case, 
and then calculating for each $x_p$ bin the average value of the double spin 
asymmetry and its variance. Again, we checked that 20 repetitions are a 
reasonable threshold to have stable numbers, since the results do not change 
significantly when increasing the repetitions from 6 to 20.

\begin{figure}[h]
\centering
\includegraphics[width=9cm]{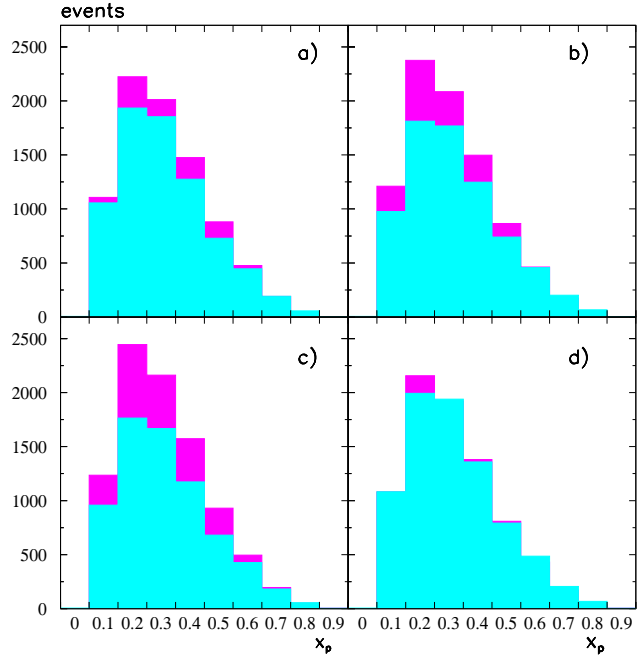}
\vspace{-0.5cm}
\caption{The sample of 17000 events for the 
$\bar{p}^\uparrow \, p^\uparrow \rightarrow \mu^+\, \mu^- \,X$ process 
where a transversely polarized antiproton beam with energy $E_{\bar{p}} = 15$ GeV 
collides on a transversely polarized proton beam with $E_p = 3.3$ GeV producing 
muon pairs of invariant mass $4 \leq M \leq 9$ GeV (for further details on the 
cutoffs, see text). a) $\langle h_1(x_p) \rangle / \langle f_1(x_p) \rangle =
\sqrt{x_p}$ (brackets mean that each flavor contribution in the numerator is 
replaced by a common average term, similarly in the denominator; for further 
details, see text). b) $\langle h_1(x_p) \rangle / \langle f_1(x_p) \rangle = 
\sqrt{1-x_p}$. c) $\langle h_1(x_p) \rangle / \langle f_1(x_p) \rangle = 1$. d) 
$\langle h_1(x_p) \rangle / \langle f_1(x_p) \rangle = 0$. For each bin, the 
darker histogram corresponds to positive values of $\cos(2\phi -
\phi_{_{S_{\bar{p}}}} - \phi_{_{S_p}})$ in Eq.~(\protect{\ref{eq:mcS}}), the 
superimposed lighter one to negative values.}
\label{fig:assia2-hist}
\end{figure}

In Fig.~(\ref{fig:assia2-hist}), the sample of 17000 "good" events for the 
$\bar{p}^\uparrow \, p^\uparrow \rightarrow \mu^+ \, \mu^- \, X$ process is
displayed in the above kinematic conditions for the collider mode. The invariant 
mass of the lepton pair is constrained in the range $4\leq M \leq 9$ GeV. The 
four panels correspond to the choices: a) $\langle h_1(x_p) \rangle / \langle 
f_1(x_p) \rangle = \sqrt{x_p}$; b) $\langle h_1(x_p) \rangle / \langle f_1(x_p) 
\rangle = \sqrt{1-x_p}$; c) $\langle h_1(x_p) \rangle / \langle f_1(x_p) \rangle 
= 1$; d) $\langle h_1(x_p) \rangle / \langle f_1(x_p) \rangle = 0$. The first two 
cases have been selected in order to have two opposite behaviours, namely 
ascending and descending, and to verify if they can be identified also in the 
corresponding asymmetries. The last two ones are displayed to set a reference 
scale for the absolute size of the asymmetry and to crosscheck the statistics 
when $A_{_{TT}}\propto \langle h_1(x_p) \rangle / \langle f_1(x_p) \rangle = 0$. 
In any case, all four choices respect the Soffer bound~\cite{Soffer:1995ww}. For 
each bin, two groups of events are stored corresponding to positive values of 
$\cos (2\phi -\phi_{_{S_{\bar{p}}}} - \phi_{_{S_p}})$ in Eq.~(\ref{eq:mcS}), 
represented by the darker histograms (events $U$), and to negative values, 
corresponding to the superimposed lighter ones (events $D$). The bins at the 
boundaries, corresponding to $x_p < 0.1$ and $x_p > 0.7$, contain a number of
events below the discussed statistical cutoffs and they will be discarded in the
corresponding asymmetry. 

\begin{figure}[h]
\centering
\includegraphics[width=9cm]{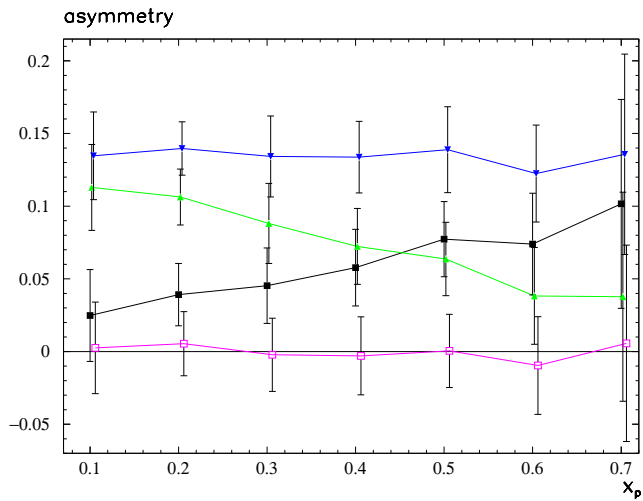}
\vspace{-0.5cm}
\caption{Asymmetry $(U-D)/(U+D)$ between cross sections in the previous figure
corresponding to darker histograms ($U$) and superimposed lighter histograms
($D$), as bins in $x_p$. Full squares for the case when 
$\langle h_1(x_p) \rangle / \langle f_1(x_p) \rangle = \sqrt{x_p}$, upward
triangles when it equals $\sqrt{1-x_p}$, downward triangles when it equals 1 and
open squares when it equals 0. Continuous lines are drawn to guide the eye. Error
bars due to statistical errors only, obtained by 20 independent repetitions of the
simulation (see text for further details).}
\label{fig:assia2-as}
\end{figure}

In Fig.~\ref{fig:assia2-as}, the corresponding double spin asymmetry $(U-D)/(U+D)$
is displayed again in $x_p$ bins for all the four choices: full squares for 
$\langle h_1(x_p) \rangle / \langle f_1(x_p) \rangle = \sqrt{x_p}$, upward
triangles for $\langle h_1(x_p) \rangle / \langle f_1(x_p) \rangle =
\sqrt{1-x_p}$, downward triangles for $\langle h_1(x_p) \rangle / \langle 
f_1(x_p) \rangle = 1$, and open squares for $\langle h_1(x_p) \rangle / \langle 
f_1(x_p) \rangle = 0$. The error bars represent statistical errors only.

From the above results, we first deduce that the considered sample of events in
the specified kinematics is sufficient to produce an average significant 
asymmetry, at most about 15\% for $\langle h_1(x_p) \rangle / \langle f_1(x_p) 
\rangle = 1$. 
When the ratio equals 0, the corresponding asymmetry consistently 
oscillates around 0 in all considered bins, also for low $x_p$ where the smaller 
error bars are due to the more dense population of events. In fact, the cross 
section is known to rapidly increase for decreasing $\tau$ and data accumulate 
in the part of the phase space corresponding to the lowest possible $\tau$ which, 
for the considered kinematics, is $\tau \gtrsim 0.08$. At the same time, in  
the range $0.1 \leq x_p < 0.4$ it seems also that the asymmetries corresponding 
to the ascending $\sqrt{x_p}$ and descending $\sqrt{1-x_p}$ functions keep this 
different behaviour. This means that, although in this limited (but significant) 
range, it should be possible to extract also information on the analytical 
dependence of the transversity upon the parton fractional momentum. For higher 
$x_p$ (and $\tau \lesssim 0.4$), the phase space is less populated and the
error bars are so large that each one of the four choices for the ratio 
$\langle h_1(x_p) \rangle / \langle f_1(x_p) \rangle$ can be confused with the
other ones. 

\begin{figure}[h]
\centering
\includegraphics[width=10cm]{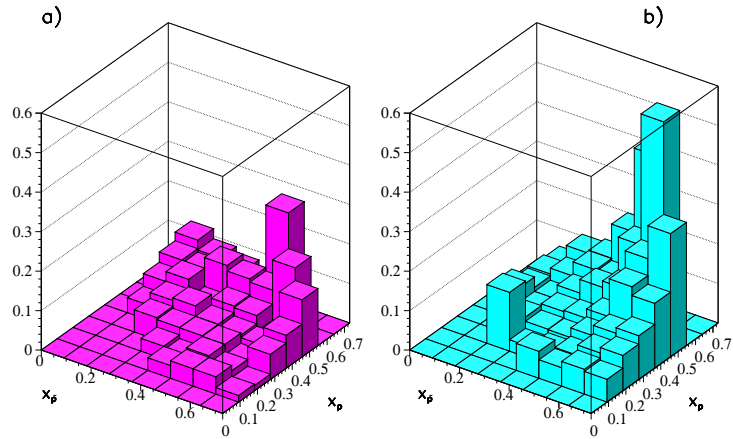}
\vspace{-0.2cm}
\caption{Unintegrated asymmetry $(U-D)/(U+D)$ for the case 
$\langle h_1(x) \rangle / \langle f_1(x) \rangle = \sqrt{x}$ in the same
conditions as the previous figure, but plotted in bins of $x_{\bar{p}}$ and $x_p$.
Left panel: distribution of average values. Right panel: distribution of the
variances, i.e. of half the statistical "error bars", obtained by 20 independent 
repetitions of the simulation (see text for further details).}
\label{fig:assia2-lego-su}
\end{figure}


This trend is confirmed and better clarified by looking at the unintegrated
asymmetry, displayed in Fig.~\ref{fig:assia2-lego-su} in bins of $x_{\bar{p}}$ 
and $x_p$ for the case $\langle h_1(x_{\bar{p}}) \rangle \, \langle h_1(x_p) 
\rangle / \langle f_1(x_{\bar{p}}) \rangle \, \langle f_1(x_p) \rangle = 
\sqrt{x_{\bar{p}}}\,\sqrt{x_p}$. Namely, it corresponds to the full squares of 
Fig.~\ref{fig:assia2-as} for the $x_{\bar{p}}$-integrated case. The left panel 
represents the bidimensional plot of the average values of the unintegrated 
double spin asymmetry, while the right panel gives the distribution of the 
variance, i.e. of half the "error bar" for each $(x_{\bar{p}},x_p)$ bin. On the
boundaries of the plot, where $x_{\bar{p}}, x_p$ are close to 1, the small
number of collected events produces large statistical errors and also 
large asymmetries due to fluctuations. Anyway, each bin containing less than 10 
events is considered not statistically relevant and the corresponding asymmetry 
and error have been artificially put to zero. Viceversa, for small $x_p \lesssim 
0.3$ the variance is small through all the $x_{\bar{p}}$ range so to give a
distinguishable integrated double spin asymmetry. Unfortunately, because of the
ascending trend of the function $\sqrt{x}$, the absolute values of the asymmetry
are small in the $x_p$ range of interest, even if statistically different from 
zero. 

\begin{figure}[h]
\centering
\includegraphics[width=10cm]{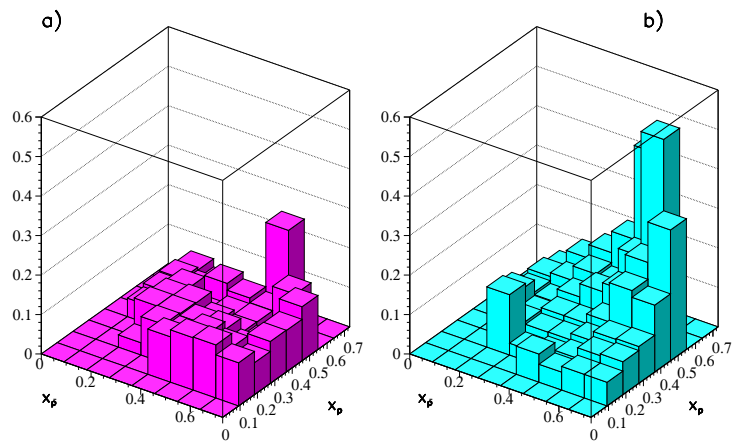}
\vspace{-0.2cm}
\caption{Unintegrated asymmetry $(U-D)/(U+D)$ for the case 
$\langle h_1(x) \rangle / \langle f_1(x) \rangle = \sqrt{1-x}$ in the same
conditions as the previous figure, plotted in bins of $x_{\bar{p}}$ and $x_p$.
Left panel: distribution of average values. Right panel: distribution of the
variances, i.e. of half the statistical "error bars", obtained by 20 independent 
repetitions of the simulation (see text for further details).}
\label{fig:assia2-lego-giu}
\end{figure}


In Fig.~\ref{fig:assia2-lego-giu} the unintegrated double spin
asymmetry is displayed for the case $\langle h_1(x_{\bar{p}}) \rangle \, \langle 
h_1(x_p) \rangle / \langle f_1(x_{\bar{p}}) \rangle \, \langle f_1(x_p) \rangle = 
\sqrt{1-x_{\bar{p}}}\,\sqrt{1-x_p}$ in the same conditions as in the previous
figure. Therefore, it corresponds to the upward triangles of
Fig.~\ref{fig:assia2-as} for the $x_{\bar{p}}$-integrated case. Similar arguments
apply to the statistical selection of the results. The only difference is that for
the relevant $x_p \lesssim 0.3$ range the absolute values of the asymmetry are
more significant because of the descending trend of the function $\sqrt{1-x}$. 
By comparing the right panel of this figure with the corresponding one in
Fig.~\ref{fig:assia2-lego-su}, we notice that errors actually do not depend on 
the selected test function. They increase with increasing $\tau$, with the 
exception of some bins that are crossed by the hyperbole $M > 4$ GeV/c.

\begin{figure}[h]
\centering
\includegraphics[width=10cm]{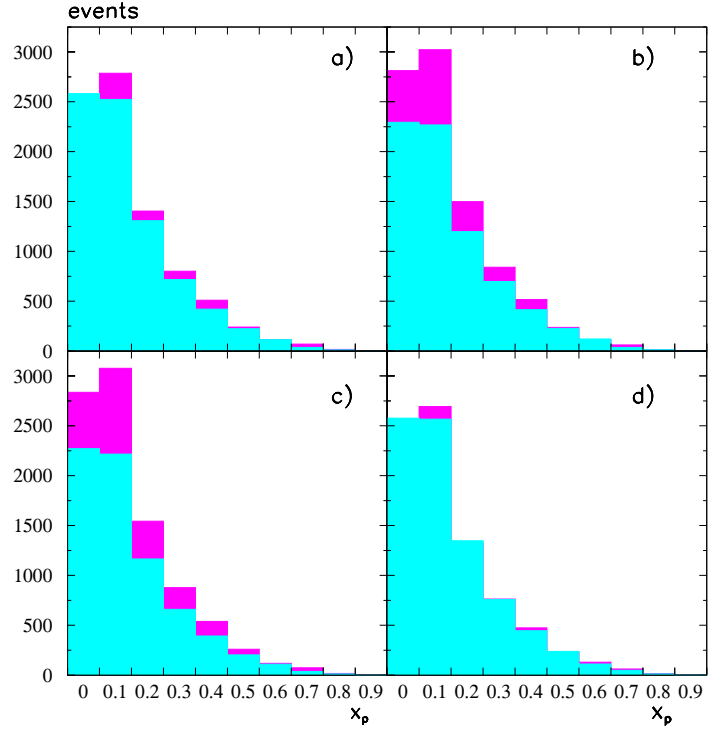}
\caption{The sample of 17000 events for the 
$\bar{p}^\uparrow \, p^\uparrow \rightarrow \mu^+\, \mu^- \,X$ process for the
lepton invariant mass $1.5 \leq M \leq 2.5$ GeV, while the other kinematic
conditions, conventions for each panel, and color codes of histograms are as in 
Fig.\protect{\ref{fig:assia2-hist}} (for further details, see text).}
\label{fig:mixed-hist}
\end{figure}


In Fig.~(\ref{fig:mixed-hist}), the sample of 17000 "good" events for the 
$\bar{p}^\uparrow \, p^\uparrow \rightarrow \mu^+ \, \mu^- \, X$ process is
displayed in the same conditions as in Fig.~\ref{fig:assia2-hist}, but with the 
invariant mass of the lepton pair in the range $1.5\leq M \leq 2.5$ GeV. 
Again, the conventions for the four panels and the color codes of the histograms
are the same as in Fig.~\ref{fig:assia2-hist}. Since the range of explored $\tau =
M^2/s$ is now 0.01-0.03, the bins for $x_p \rightarrow 0$ are more 
populated than before, while for $x_p > 0.7$ yet they are statistically not 
significant and again they will be discarded in the asymmetry plot.

\begin{figure}[h]
\centering
\includegraphics[width=10cm]{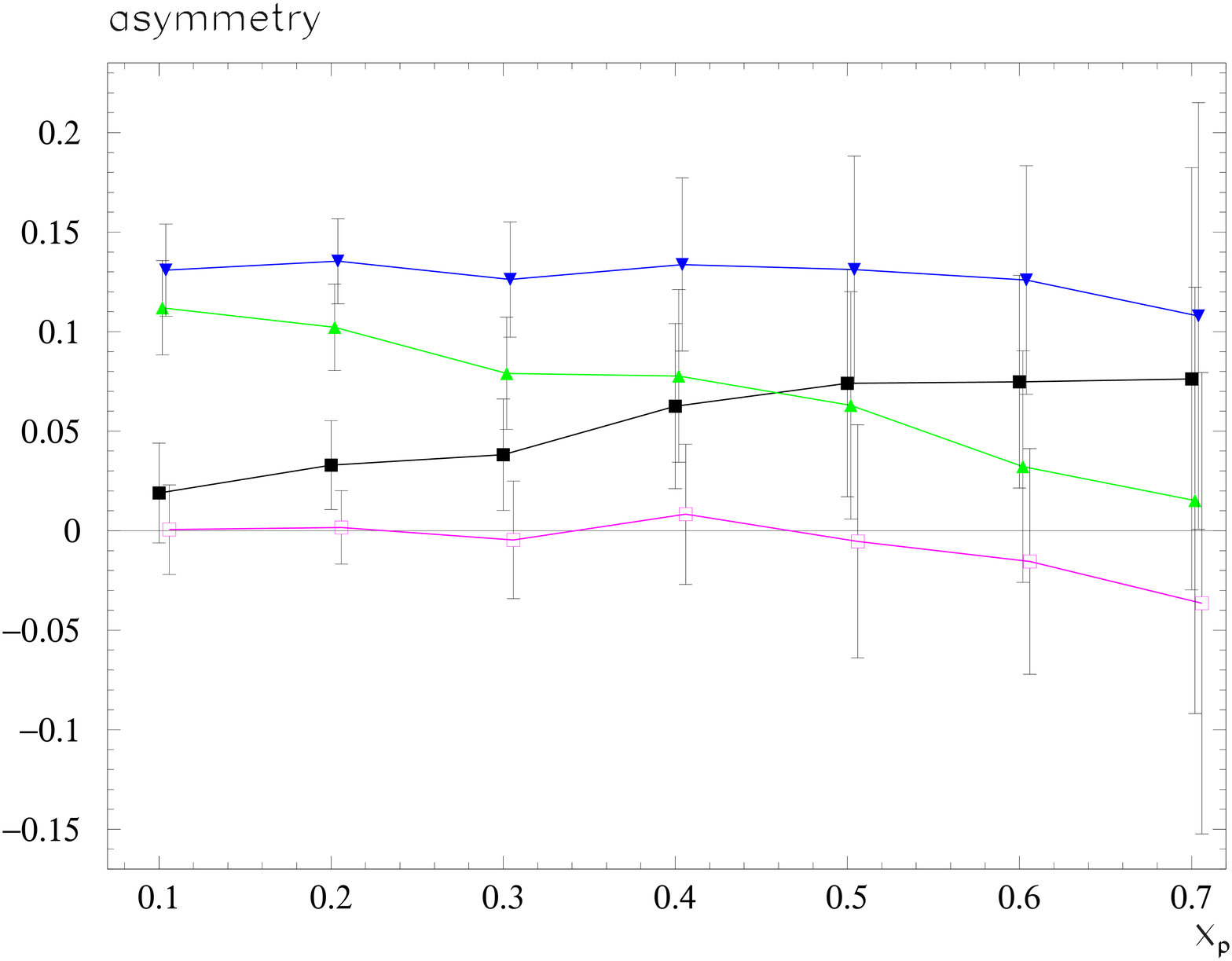}
\caption{Double spin asymmetry $(U-D)/(U+D)$ between cross sections in the 
previous figure corresponding to darker histograms ($U$) and superimposed lighter 
histograms ($D$), as bins in $x_p$. Notations as in 
Fig.~\protect{\ref{fig:assia2-as}}. Continuous lines are drawn to guide the eye. 
Error bars due to statistical errors only, obtained by 20 independent repetitions 
of the simulation (see text for further details).}
\label{fig:mixed-as}
\end{figure}


In Fig.~\ref{fig:mixed-as}, the corresponding double spin asymmetry $(U-D)/(U+D)$
is displayed with the same conventions as in Fig.~\ref{fig:assia2-as}. The larger
population for small $x_p$ bins reflects in smaller error bars for all four
choices of the ratio $\langle h_1(x_p) \rangle / \langle f_1(x_p) \rangle$.
Consequently, in the range $0.1 \lesssim x_p \lesssim 0.3$ the functions
$\sqrt{x_p}$ (full squares) and $\sqrt{1-x_p}$ (upward triangles) are even more 
distinguishable than in the previous case. 

\begin{figure}[h]
\centering
\includegraphics[width=9cm]{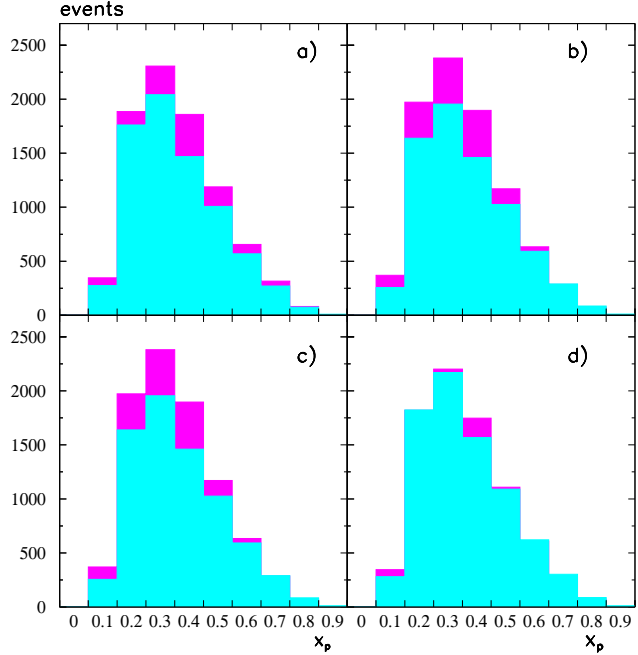}
\caption{The sample of 17000 events for the 
$\bar{p}^\uparrow \, p^\uparrow \rightarrow \mu^+\, \mu^- \,X$ process where a 
transversely polarized antiproton beam with energy $E_{\bar{p}} = 15$ GeV hits a
transversely polarized proton target producing muon pairs with invariant mass 
$1.5 \leq M \leq 2.5$ GeV and $s=30$ GeV$^2$. Conventions for each panel and 
color codes of histograms are as in Fig.\protect{\ref{fig:assia2-hist}} (for 
further details, see text).}
\label{fig:panda-hist}
\end{figure}

As already recalled in the previous section, we have also explored the typical 
kinematics suggested by the PANDA collaboration in its proposal at HESR at 
GSI~\cite{panda}. Namely, the operational mode where the antiproton beam with 
energy $E_{\bar{p}}=15$ GeV hits a proton target and produces lepton pairs with 
low invariant masses; we have considered the range $1.5\leq M \leq 2.5$ GeV for 
the same reasons as above. We have still assumed that the transverse 
polarization of both beam and target is 50\% on the average, such that the
corresponding overall dilution factor is 0.25. This corresponds to take 
$|{\bf S}_{_{Tp}}| = |{\bf S}_{_{T\bar{p}}}| = 1$ in Eqs.~(\ref{eq:mcS}) and 
(\ref{eq:mcatt}) for the 25\% of the events, while taking them 0 for the remaining
75\%. The resulting cm square energy is $s=30$ GeV$^2$ and the range 
of explored $\tau$ is approximately 0.07-0.2. Consequently, the integrated event 
distribution of Fig.~\ref{fig:panda-hist} is more populated at higher $x_p$ bins 
for all the four $\langle h_1 \rangle / \langle f_1 \rangle$ ratios explored 
(notations, conventions and histogram color codes are as in 
Fig.~\ref{fig:assia2-hist}).

\begin{figure}[h]
\centering
\includegraphics[width=9cm]{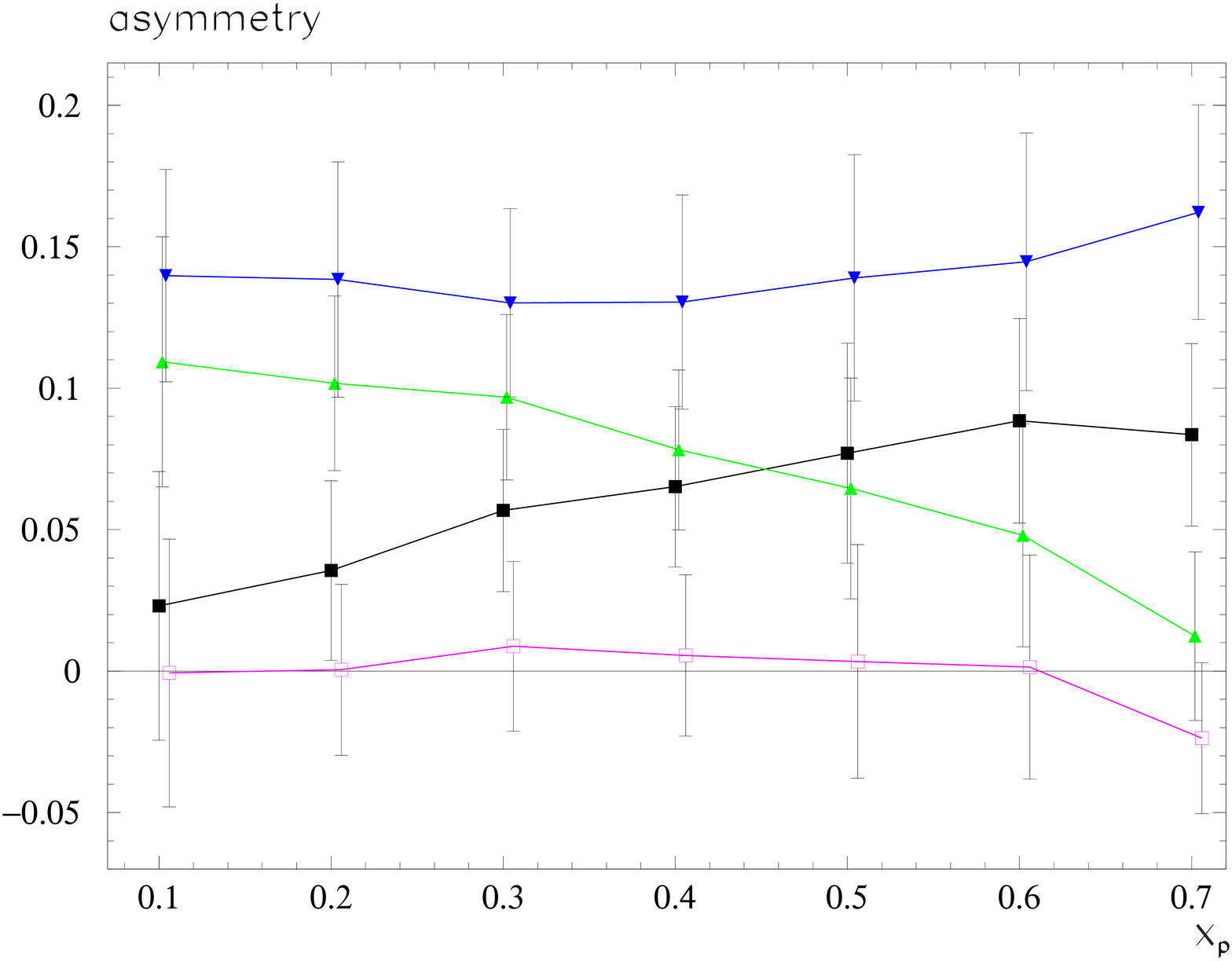}
\caption{Double spin asymmetry $(U-D)/(U+D)$ between cross sections in the 
previous figure corresponding to darker histograms ($U$) and superimposed lighter 
histograms ($D$), as bins in $x_p$. Notations as in 
Fig.~\protect{\ref{fig:assia2-as}}. Continuous lines are drawn to guide the eye. 
Error bars due to statistical errors only, obtained by 20 independent repetitions 
of the simulation (see text for further details).}
\label{fig:panda-as}
\end{figure}


In Fig.~\ref{fig:panda-as}, the double spin asymmetry $(U-D)/(U+D)$, corresponding
to cross sections of the previous figure, is displayed with the same conventions 
as in Fig.~\ref{fig:assia2-as}. At variance with the result of
Fig.~\ref{fig:mixed-as}, the explored portion of phase space for larger $\tau$
(hence, for larger $x_p$ bins) is not favoured by the given $1/\tau$ qualitative
dependence of the cross section: a lower density of events reflects in larger 
statistical error bars which prevent from clearly distinguishing each one of the
four considered forms for the $\langle h_1 \rangle / \langle f_1 \rangle$ ratio.
This means that for the considered sample of 17000 "good" events, it is not yet
clear how to extract information on the analytical structure of $h_1(x)$. 
However, it should be possible at least to observe a nonvanishing asymmetry and 
give an estimate of its magnitude and sign. 

The relevant message of previous figures is that it is crucial to consider
integrated distributions in one parton fractional momentum only, in order to 
reasonably populate bins and to reach a deconvolution of transversity from the 
product $h(x_{\bar{p}}) h(x_p)$. The price to pay is that the useful phase space
in the other fractional momentum is reduced to few bins. It is easy to estimate
where the maximum of the distribution is located. Assuming that the bidimensional 
$(x_{\bar{p}}, x_p)$ distribution is dominated by the $1/\tau$ factor associated 
with the elementary $q \bar{q}$ fusion into a virtual photon, the distribution 
is $1/(x_{\bar{p}} x_p)$ for $\tau > M^2_{min}/s$, and 0 otherwise. Consequently, 
the $x_{\bar{p}}$-integrated distribution has the form $\log 
(x_p s/M^2_{min})/x_p$ and reaches its peak value for $\log (x_p s/M^2_{min}) = 
1$, i.e. for $x_p = e M_{min}^2/s \approx 3 M_{min}^2/s$. For $M = 4$ GeV and 
$s = 200$ GeV$^2$, $x_p \approx 0.22$ in agreement with Fig.~\ref{fig:assia2-as}. 
It is evident that it is possible to span the domain of valence contributions for 
$s$ in the range 100-300 GeV$^2$ and $M \gtrsim 4$ GeV. A similar conclusion holds
by directly considering the unintegrated distribution, where the relevant
contribution of annihilating valence (anti)partons corresponds to $x_{\bar{p}} =
x_p \approx 0.3$, such that $\tau = x_p x_{\bar{p}} \approx 0.1$ can be reached
again for $100 \lesssim s \lesssim 300$ GeV$^2$ and $M\gtrsim 4$ GeV. 

Moreover, exploring different masses at the same $s$ (as we did in
Figs.~\ref{fig:assia2-as} and \ref{fig:mixed-as}) is useful to estimate the 
role of higher twist effects, since the latter can be classified according to 
powers of $M/M_p$, where $M_p$ is the proton mass. However, our results do not 
include these corrections since precision calculations are beyond the scope of 
the present work.

In conclusion, it seems that in the collider mode for the HESR at GSI, where the 
cm square energy is $s=200$ GeV$^2$, a sample of 17000 Drell-Yan events from 
collisions of transversely polarized antiproton and proton beams with proper 
lepton invariant masses is sufficient to generate sizeable double spin asymmetries
from which information can be extracted about the functional dependence of the
transversity, but limited to the range $0.1 \lesssim x \lesssim 0.3$. For the
lower case $s=30$ GeV$^2$ in the fixed-target mode, the double spin asymmetry is 
still sizeable, but the larger statistical error bars do not allow for such a 
clean extraction.


\section{Conclusions}
\label{sec:end}

In a previous paper~\cite{Bianconi:2004wu}, we produced a Monte Carlo simulation
to study the physics case of a Drell-Yan process with unpolarized antiproton 
beams and transversely polarized proton targets. Here, we have considered the 
fully polarized $\bar{p}^\uparrow \, p^\uparrow \rightarrow \mu^+\,\mu^-\,X$ 
process in order to explore the leading transverse spin structure of the nucleon. 
Both works are finalized at the kinematics of the High Energy Storage Ring (HESR),
a source for (polarized) antiprotons under development at GSI. In fact, using
antiproton beams offers the advantage of involving unsuppressed distributions of
valence partons: the transversely polarized quark in a transversely polarized
proton, and the transversely polarized antiquark in a transversely polarized
antiproton. 

Different kinematical options have been considered. Since the cross section 
fastly decreases for increasing $\tau = M^2/s = x_p x_{\bar{p}}$ (where $M$ is 
the lepton pair invariant mass and $s$ is the center-of-mass square energy), 
events tend to accumulate in the phase space part corresponding to the smallest 
$\tau$ allowed by the mass cutoff, which is dominated by the valence contribution
to parton distributions. It seems convenient to reach such low values of $\tau$ by
adequately increasing $s$. The $x_{\bar{p}}$-integrated distributions present a 
Poisson-like shape in the parton fractional momentum $x_p$ with a peak at 
$x_p \approx 3 M^2_{min}/s$ and width 0.1-0.2. Outside this range, the event 
distributions are of difficult analysis. For typical $M_{min}$ values, the most 
interesting $s$ range turns out to be 100-300 GeV$^2$, where the possibility of 
repeating the experiment at different $s$ values allows to analyze different 
$x_p$ ranges. 

We have simulated the fully polarized Drell-Yan process using antiproton beams 
with energy $E_{\bar{p}}=15$ GeV and (transverse) polarization 50\%, and proton 
beams with the same polarization and energy $E_p=3.3$ GeV, such that $s=200$ 
GeV$^2$. To avoid the $c\bar{c}$ threshold and other resonances in the mass
spectrum of the lepton pair, we selected the invariant mass $M$ in the ranges 
4-9 GeV and 1.5-2.5 GeV. The corresponding explored $\tau$ ranges are 0.08-0.4 
and 0.01-0.03, respectively. We have also explored the case where the transversely
polarized antiproton beam of 15 GeV hits a transversely polarized fixed proton 
target producing lepton pairs with invariant mass in the same range 1.5-2.5 GeV. 
In this case, $s=30$ GeV$^2$ and the $\tau$ range is approximately 0.07-0.2. The 
mass range 4-9 GeV has not been considered here, because it implies average $x_p$ 
values where the parton model underlying the simulation is not appropriate. 

In all cases, the transverse momentum of the dimuon couple has been limited to 
$q_{_T} > 1$ GeV/$c$ and its zenithal-angle distribution is restricted to the 
range 60-120 deg. The former cut is induced by the need to avoid complicated soft
mechanisms and to match the experimental requirements of the collider setup. The 
latter cut prevents the angular dependence of the cross section from diluting the 
related asymmetry. All these cuts produce a remarkable reduction of the Drell-Yan 
events: the considered initial sample of 80000 events (only affected by the mass 
cut) is reduced to 17000. 
 
At leading twist in the cross section, the contribution of interest has a 
characteristic azimuthal dependence of the kind $\cos (2\phi - 
\phi_{_{S_{\bar{p}}}} - \phi_{_{S_p}})$, where $\phi$ is the azimuthal angle of 
the final lepton pair and $\phi_{_{S_{p(\bar{p})}}}$ is the azimuthal position of 
the (anti)proton transverse spin in the Collins-Soper frame. Hence, for each 
randomly distributed $\phi_{_{S_{\bar{p}}}}$ events have been accumulated for 
parallel $(\phi_{_{S_{\bar{p}}}} = \phi_{_{S_p}})$ and antiparallel 
$(\phi_{_{S_{\bar{p}}}} = \phi_{_{S_p}} + \pi)$ beam polarizations. The 
corresponding azimuthal asymmetry has been built either integrating upon all 
variables but the quark fractional momentum $x_p$, or keeping also the 
corresponding antiquark $x_{\bar{p}}$ dependence and building bidimensional plots.

For each kinematical case, we have considered four functional dependences for
the transversity, each one with a very peculiar trend, namely constant, 
ascending, descending, and zero, but all satisfying the Soffer bound. The goal 
is to recover the same different trend also in the corresponding double spin 
asymmetry, which means that feasibility conditions for an unambiguous extraction 
of the transversity could be established. In the collider mode with $s=200$ 
GeV$^2$, the double spin asymmetry has very small statistical error bars for 
$0.1\lesssim x_p \lesssim 0.3$. With this limitation, it seems that a sample of 
17000 events satisfying all the described cutoffs, should be sufficient to grant 
the identification of the analytical behaviour of the transversity. This statement
is still valid at the lower $M$ range considered by keeping the same $s$, because 
the resulting $\tau$ range is shifted to lower values and the statistics is 
higher. Viceversa, for the option where the antiproton beam hits a fixed proton 
target, a lower $s$ results in larger error bars: measuring a nonvanishing double
spin asymmetry seems possible, but the extraction of the transversity looks more
problematic. 

In conclusion, with the present simulation we have explored the feasibility
conditions for an unambiguous extraction of the transversity from Drell-Yan data
with polarized antiproton beams; we hope to have contributed to the studies of the
physics case for hadronic collisions with antiproton beams at the HESR at GSI.


\bibliographystyle{apsrev}
\bibliography{hadron}


\end{document}